\documentclass[journal=jpcbfk,manuscript=article]{achemso}
\usepackage{textcomp}
\usepackage{longtable}
\usepackage{multirow}
\usepackage{amsmath}
\usepackage{subfigure}
\usepackage{bm} 

\usepackage[usenames,dvipsnames]{color}
\usepackage{soul, xcolor}
\usepackage{threeparttable}
\usepackage[normalem]{ulem}

\usepackage{xr}
\SectionNumbersOn

\author{Marco Pezzella, Debasish Koner} \affiliation{Department of Chemistry,
  University of Basel, Klingelbergstrasse 80, CH-4056 Basel,
  Switzerland.}

\author{Markus Meuwly} \affiliation{Department of Chemistry,
  University of Basel, Klingelbergstrasse 80, CH-4056 Basel,
  Switzerland.}
\email{m.meuwly@unibas.ch}

\title{Formation and Stabilization of Ground and Excited State Singlet
  O$_2$ upon Recombination of $^3$P Oxygen on Amorphous Solid Water}

\begin{document}

\date{\today}

\begin{abstract}
The recombination dynamics of $^3$P oxygen atoms on cold amorphous
solid water to form triplet and singlet molecular oxygen (O$_2$) is
followed under conditions representative for cold clouds. It is found
that both, formation of ground state ($X ^3 \Sigma_{g}^{-}$) O$_2$ and
molecular oxygen in the two lowest singlet states ($a ^1\Delta_g$ and
$b ^1\Sigma_g^+$) is possible and that the species can stabilize. The
relative proportions of the species is approximately 1:1:1. These
results also agree qualitatively with a kinetic model based on
simplified wavepacket simulations. As the chemical reactivity of
triplet and singlet O$_2$ is different it is likely that substantial
amounts of $a ^1\Delta_g$ and $b ^1\Sigma_g^+$ oxygen influences the
chemical evolution of cold clouds.
\end{abstract}

\section{Introduction}
\label{sec:intro}
The role of electronically excited states of oxygen, in particular
that of singlet O$_2$, is well established in biological,
atmospheric\cite{parish:2012,hase:2019} and combustion
chemistry\cite{Chukalovsky:2010,starik:2015,Chukalovsky:2016}. Combustion
processes involving H$_2$, CO or CH$_4$ are accelerated in the
presence of O$_2$ in its $a ^1\Delta_g$ and $b ^1\Sigma_g^+$ states,
compared to reactions with O$_2$ in its $X ^3\Sigma_g^-$ ground
state.\cite{starik:2015} No such effects are observed for the
combustion of N$_2$\cite{starik:2015}. Computational
studies\cite{hase:2019} on CH$_2$ + O$_2$ and thiophene
\cite{parish:2012} probe different aspects of O$_2$ excited state
reactivity: in the first case no difference between the ground state
and the second excited state is encountered, while for thiophene the
singlet oxygen channel seems to be dominant.\\

\noindent
Recently, the possibility of oxygen diffusion\cite{MM.oxy:2018} and
recombination of two $^3$P oxygen to form ground state O$_2$ in dark
molecular clouds was established.\cite{MM.oxy:2019} Given this, it is
also of interest to explore the possibility that molecular oxygen can
be formed in low lying electronically excited states under
interstellar conditions. Experimental spectroscopic observations show
that recombination of two oxygen atoms ($^3P$) generated from
photolysis of $^{16}$O$_2$ using far-ultraviolet light in neon
matrices at low temperature ($T \sim 5$ K) leads to formation of O$_2$
in its $X ^3\Sigma_g^-$, $a ^1\Delta_g$, $b ^1\Sigma_g^+$, and
additional, more highly excited electronic states\cite{ogilvie:2018}
although the relative populations of the states were not reported.
The first ($^1\Delta_g$) and second ($^1\Sigma_g^+$) excited states
are of particular interest, due to their higher reactivity compared to
the ground state.  In the gas phase and in isolation the two
transitions $a^1\Delta_g \rightarrow$ $X^3\Sigma_g^-$ and $b^1
\Sigma_g^+ \rightarrow$ $X^3\Sigma_g^-$ are symmetry forbidden with
radiative lifetimes of 4000 and 150 s, induced by magnetic-dipole and
electric-quadrupole interactions,\cite{parker:2014} respectively. A
major contributor to the $b^1\Sigma_g^+ \rightarrow$ $X^3\Sigma_g^-$
transition is the first order Spin-Orbit coupling close to the
$(\nu=28, N=5)$ of the ground state.\cite{wodtke:1999} Collision
induced emission has been reported to accelerate the $a^1\Delta_g
\rightarrow$ $X^3\Sigma_g^-$ transition, being 9 orders of magnitude
faster (500 $\mu$s vs. 4000 s) than the radiative
emission.\cite{Hidemori:2012} However, in the presence of an
environment, these transitions can occur due to the perturbations
induced by the environment. Relaxation from the $^1\Sigma_g^+$ state
to $^3\Sigma_g^-$ occurs \textit{via} Inter System Crossing (ISC),
governed by Spin Orbit Coupling (SOC) that can be described using the
Landau Zener (LZ) formalism.\cite{minaev2003,dayou2005} \\

\noindent
Here, the possibility is explored that upon O($^3$P) + O($^3$P)
recombination on Amorphous Solid Water (ASW) not only the ground
($X^3\Sigma_g^-$), but also electronically excited states of molecular
oxygen, i.e. O$_2$ ($b^1\Delta_g$ and $a^1\Sigma_g^+$) are formed,
stabilized and populated. ASW, which is a form of glassy water, is
considered to be the main component of ices on top of the small grains
present in interstellar clouds.\cite{angell:2004,burke:2010} The high
porosity of ASW\cite{bossa:2014,bossa:2015,cazauxs:2015} makes it a
good catalyst for gas-surface reactions involving
oxygen\cite{Ioppolo:2011,romanzin:2011,Chaabouni:2012,oxy.diff.minissale:2013,o2.dulieu:2016,MM.oxy:2018,MM.oxy:2019},
hydrogen\cite{hama:2013},
carbonaceous\cite{co.form.minissale:2013,minissale:2016} and
nitrogen-containing\cite{no1.minissale:2014} species and helps
maintaining those species on the
surface.\cite{dulieu:2016,minissale:2018} Using reactive molecular
dynamics simulations\cite{msarmd} with nonadiabatic transitions the
dynamics, relaxation and population distribution after partial
vibrational equilibration of O$_2$ in the three lowest electronic
states is characterized in the following.\\

\section{Results}
In the following, a two- and a three-state model is explored. The
two-state model provides an overview of the expected dynamics for an
electronic transition which becomes allowed in the presence of an
environment. For the more complete three-state model only two out of
the three transitions occur.\\

\begin{figure}
\centering \includegraphics[scale=0.30]{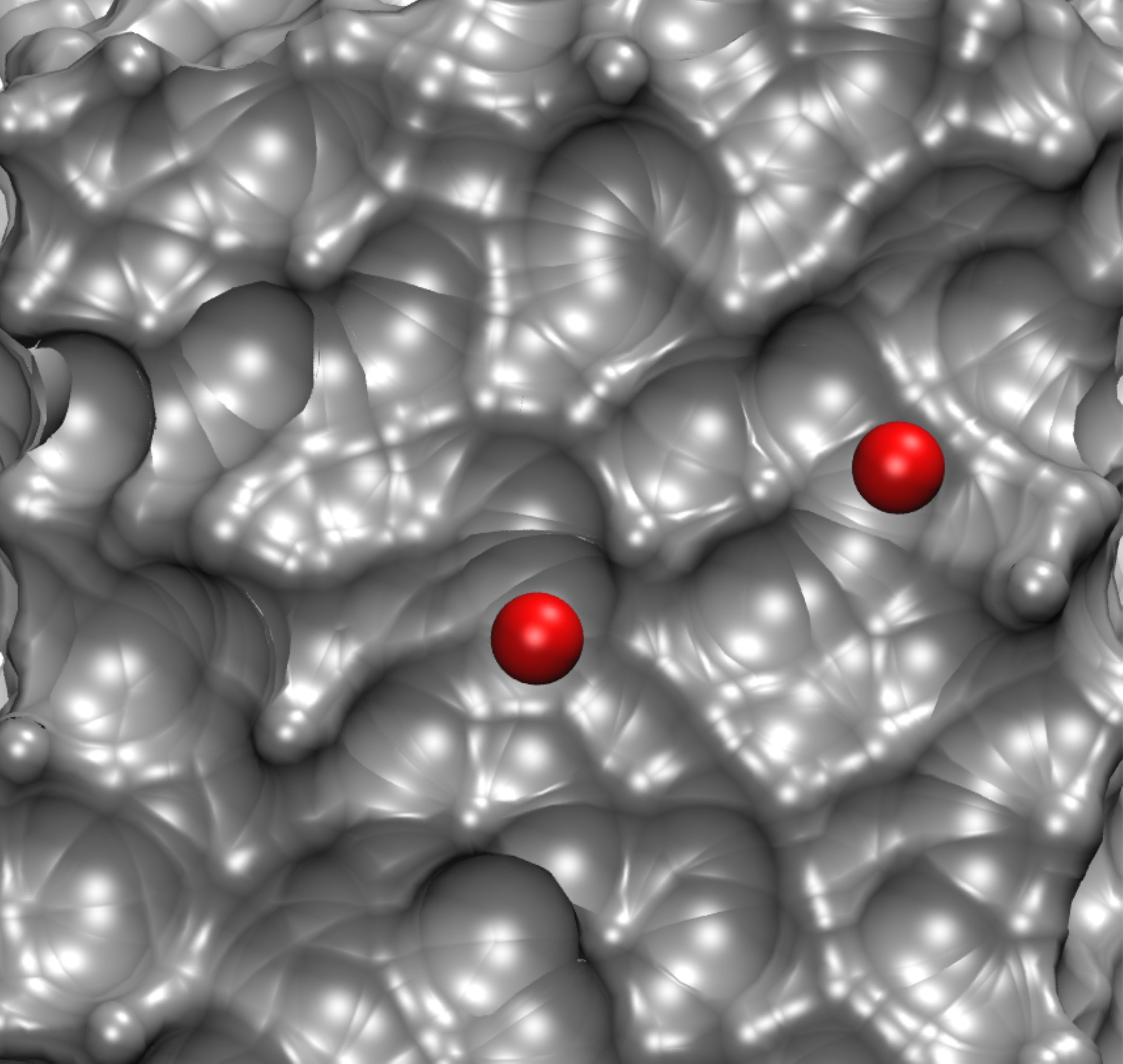}
\caption{The simulation system with water bulk represented in grey,
  with the two oxygen atoms on top of the surface in red.}
\label{fig:fig1}
\end{figure}

\subsection{Two state model}
First, a two state model involving the $X^3 \Sigma_g^-$ and $b^1
\Sigma_g^+$ states is considered. For this, different types of
simulations were carried out: a) 2 ns simulations with both oxygen
atoms inside bulk ASW; b) five 8 ns simulations with the two oxygen
atoms on the surface and rebinding into the $^3\Sigma_g^-$ state
initially; c) 2200 simulations run for 400 ps on the surface (1100
recombining into the $^3\Sigma_g^-$ state and 1100 into the
$^1\Sigma_g^+$ state).\\

\noindent
For the analysis two quantities are considered: the time between two
consecutive transitions $\tau_{\rm c}$ and the total crossing time
from the first to the last transition, $\tau_{\rm LZ}$. Per
definition, after $\tau_{\rm LZ}$ no further transitions between the
states are observed.\\

\noindent
For simulations inside ASW bulk (see Figure S2), O$_2$
is formed in its $b ^1\Sigma_g^+$ state after $\sim 750$ ps and
stabilized after two scattering events at 103.5 and 670 ps (see Figure
S2), with $\tau_{\rm LZ}=30$ ps. This time interval is
characterized by the abrupt change in the kinetic energy by $\sim 40$
kcal/mol due to the difference in potential energy between the two
states, see Figure S1. After this time no further
transitions are encountered during the simulation and the molecule
vibrationally relaxes and is stabilized in the $b^1\Sigma_g^+$
state.\\

\begin{figure}
\centering
\includegraphics[scale=0.40]{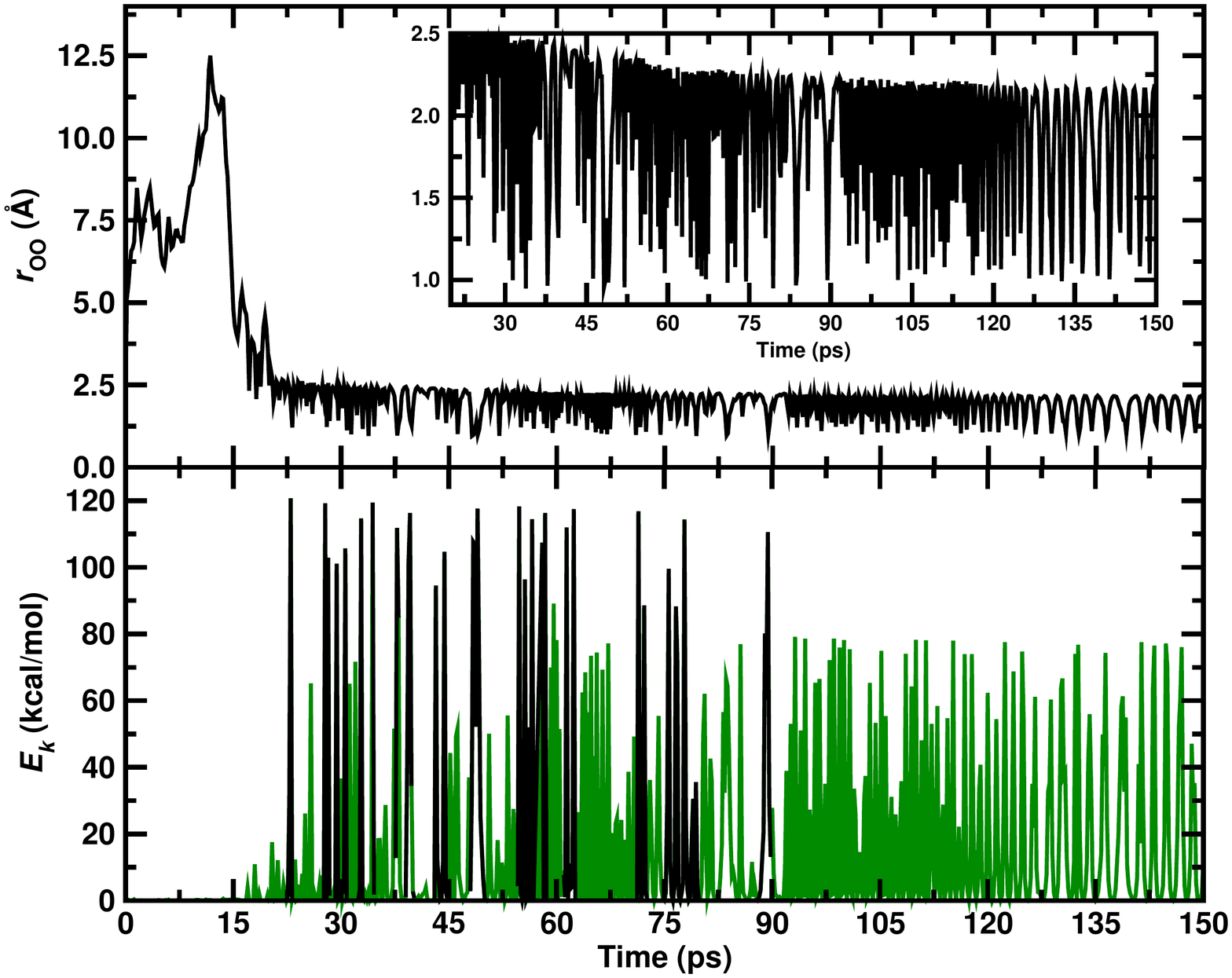} 
\caption{Top: time evolution of the interatomic distance between two
  oxygen atoms on top of ASW. Recombination occurs after 16
  ps. Bottom: kinetic energy for O$_2$ system during 150 ps
  . Formation of the bound state is reflected by the sharp increase of
  the kinetic energy of the two oxygen atoms at 16 ps. Transitions
  between the $X^3\Sigma_g^-$ (black) $b^1\Sigma_g^+$ (green) are
  observed between 17 and 100 ps. After this the system localizes on
  the $b^1\Sigma_g^+$ state. The average over the green and black
  traces also reflects the different binding energy (38.7 kcal/mol)
  for the two electronic states, see Figure S1.}
\label{mfig:ts100}
\end{figure}

\noindent
Results for a simulation on the ASW surface are shown in Figure
\ref{mfig:ts100}. Starting from an initial separation of 4.8 \AA\/,
recombination occurs after 20 ps with the molecule forming in the $b
^1\Sigma_g^+$ state followed by an extended time ($\tau_{\rm LZ}=70$
ps) during which crossings between the two states occur with final
relaxation in the $b ^1\Sigma_g^+$ state. The identity of the state is
followed explicitly in the simulations. This also allows to trace the
kinetic energy of the O$_2$ molecule during the time it samples one or
the other state, see green and black traces in Figure
\ref{mfig:ts100}.\\

\noindent
Five simulations were run by rebinding into the $b^1\Sigma_g^+$ state
for a total of 8 ns to determine whether further transitions are
observed after the molecule relaxes below the crossing point after
$\tau_{\rm LZ}$. Three trajectories lead to bound O$_2$: one in the
$X^3\Sigma_g^-$ and the other two in the $b^1\Sigma_g^+$ state and
transitions occur with a sharp distribution peaked at the crossing
point (2.209 \AA\/). In the other two simulations the two atoms do not
recombine within 8 ns but rather continue to sample an unbound
state. The time interval between the first and last transition
(i.e. $\tau_{\rm LZ}$) and between two single transitions
(i.e. $\tau_c$) is consistent with results in Figure
S4 discussed further below. The time series of two of
those simulations are reported in Figure S3.\\

\noindent
Next, the results from the 2200 rebinding simulations on the surface
are analysed. Initially, the two O atoms are separated by $6.07 \pm
2.13$ \AA\/. For $\sim 80$ \% of the simulations O$_2$ is formed,
consistent with previous work.\cite{MM.oxy:2018} Half of these
simulations initially recombine into the $X^3\Sigma_g^-$ state and the
other half into $b^1\Sigma_g^+$. The average time interval between two
consecutive transitions ($\langle\tau_{\rm c}\rangle$) is $47.4 \pm
11.7$ fs, for an average number of $1224 \pm 537$ transitions per
trajectory, independent of the initial state into which rebinding
takes place. The distribution $p(\tau_{\rm c})$ is shown in Figure
S4. On average, one crossing every two vibrational
periods is observed before sufficient vibrational energy has been
dissipated and the crossing point can not be reached anymore
energetically. After this, vibrational relaxation on the final
electronic state takes place on considerably longer time scales.\\

\begin{figure}
\centering \includegraphics[scale=0.30]{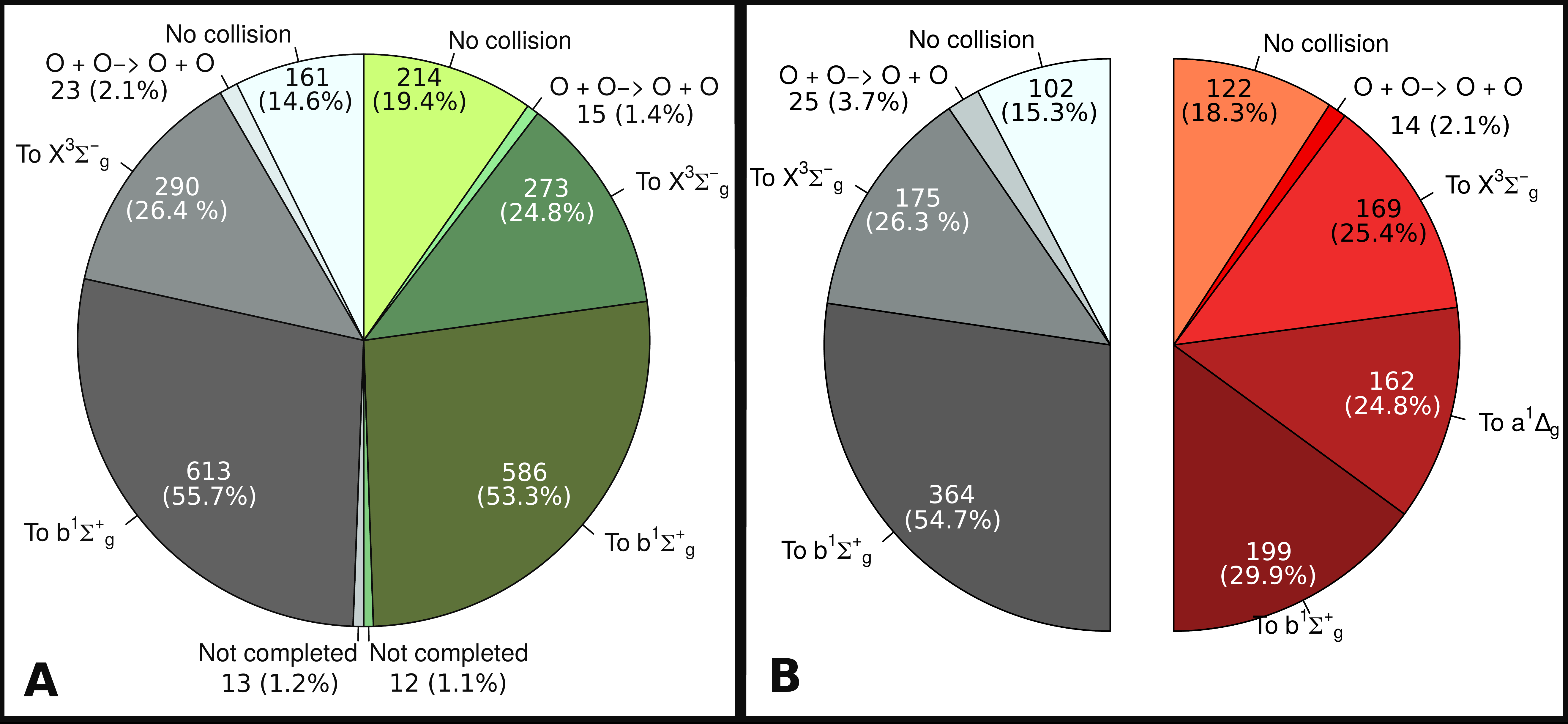}
\caption{Panel A: Classification of outcome of the simulations, in
  black those rebinding into the $X ^3\Sigma_g^-$ state and in green
  those initially rebinding into the $b ^1\Sigma_g^+$ state. Within
  400 ps, 82.1$\%$ of them lead to the O$_2$ recombination. In both
  sets, the majority of simulations leads to the $b^1\Sigma_g^+$,
  while $\sim \frac{1}{3}$ reaches the ground state.  Panel B:
  comparison between the two state model (left, grey) and the
  three-state model (right, red). For the two-state model, 666 random
  simulations from the two states models were sampled. The population
  of the $X^3\Sigma_g^-$ is similar for the two models whereas the
  population of the $b^1\Sigma_g^+$ state from the two-state model
  splits into two similar fractions for the three-state model which
  includes two excited states, $b^1\Sigma_g^+$ and $a^1\Delta_g$. The
  remaining channels are similarly populated.}
\label{fig:diagrams}
\end{figure}

\noindent
Out of the 1100 trajectories which recombine into one of the two
electronic states, 903 and 859 simulations localize either in $X^3
\Sigma_g^-$ or $b^1 \Sigma_g^+$, see Figure
\ref{fig:diagrams}. Following this, O$_2$ vibrational relaxation takes
place on considerably longer time scales. The distribution of O--O
separations $r$ at which changes in the electronic state occur is
shown in Figure S5A. From all 2200 trajectories, $\sim
66$ \% (1756 trajectories) recombine into the $^1\Sigma_g^+$ state
while 33 \% lead to the electronic ground state $X^3
\Sigma_g^-$. These fractions are independent of the initial condition,
i.e. whether initially recombination into the $X^3 \Sigma_g^-$ or $b^1
\Sigma_g^+$ state occurs, which indicates that the simulations are
converged.\\

\noindent
Of all simulations, a fraction of 14 \% and 19 \% for the $X^3
\Sigma_g^-$ and $b^1 \Sigma_g^+$ states, respectively, does not lead
to recombination and stabilization of O$_2$. Instead, the two oxygen
atoms remain separated on the surface, see Figure
\ref{fig:diagrams}A. For a small number of trajectories (2\% and 1\%,
respectively) a single collision leads to O$_2$ with subsequent
scattering and dissociation back into two separated oxygen
atoms. Finally, there is also a small number of trajectories (13 which
initially recombine into $X^3 \Sigma_g^-$ and 12 that recombine into
$b^1 \Sigma_g^+$) which have not settled into a final electronic state
after 400 ps and exploration of the electronic manifold continues on
longer time scales.\\

\noindent
The two state model indicates that recombination of two $^3$P oxygen
atoms into O$_2$ in both electronic states is possible. Furthermore,
vibrational relaxation and stabilization in these two states occurs on
considerably longer time scales than a few hundred picoseconds, as was
already found in previous work.\cite{MM.oxy:2019} For a more
comprehensive characterization, the third electronic state ($a
^1\Delta_g$), that also correlates with $^3$P oxygen is also included
in a next step.\\

\subsection{Three state model}
As a more realistic scenario, a three state model that includes the
first three electronic states ($X^3\Sigma_g^-$, $a^1\Delta_g$ and
$b^1\Sigma_g^+$) is considered. The transition between the
$X^3\Sigma_g^-$, and the $b^1\Sigma_g^+$ state is treated in the same
way as for the 2-state model. For transitions between $X^3\Sigma_g^-$
and $a^1\Delta_g$ it is noted that the \textit{ab initio} calculations
show that the spin orbit coupling is different from zero only in the
coupling region (see Figure S7) whereas no transitions
between the $a^1\Delta_g$ and $b^1\Sigma_g^+$ states are considered
because the two potential energy curves do not cross and SOC and NAC
matrix elements are zero. Overall, the $^3\Sigma_g^-$
$\longleftrightarrow$ $^1\Delta_g$ and $^3\Sigma_g^-$
$\longleftrightarrow$ $^1\Sigma_g^+$ are included in this model.\\

\noindent
For the three-state model 666 simulations were run.  While O$_2$ is in
one of the two excited singlet states, the only possible transition
leads to the ground state. This reflects the fact that the SOC and NAC
matrix elements between the two excited states is zero. For a
transition from the ground state to both excited states the more
probable of the two is chosen. To determine which transition takes
place, $P_{\rm LZ}^{j \rightarrow k}$ is evaluated for both
transitions and compared with a random number ($z$). If $P_{\rm LZ}^{j
  \rightarrow k} > z$ for both transitions, the one with the larger
value of $P_{\rm LZ}^{j \rightarrow k}$ is chosen.\\

\begin{figure}
\centering \includegraphics[scale=0.125]{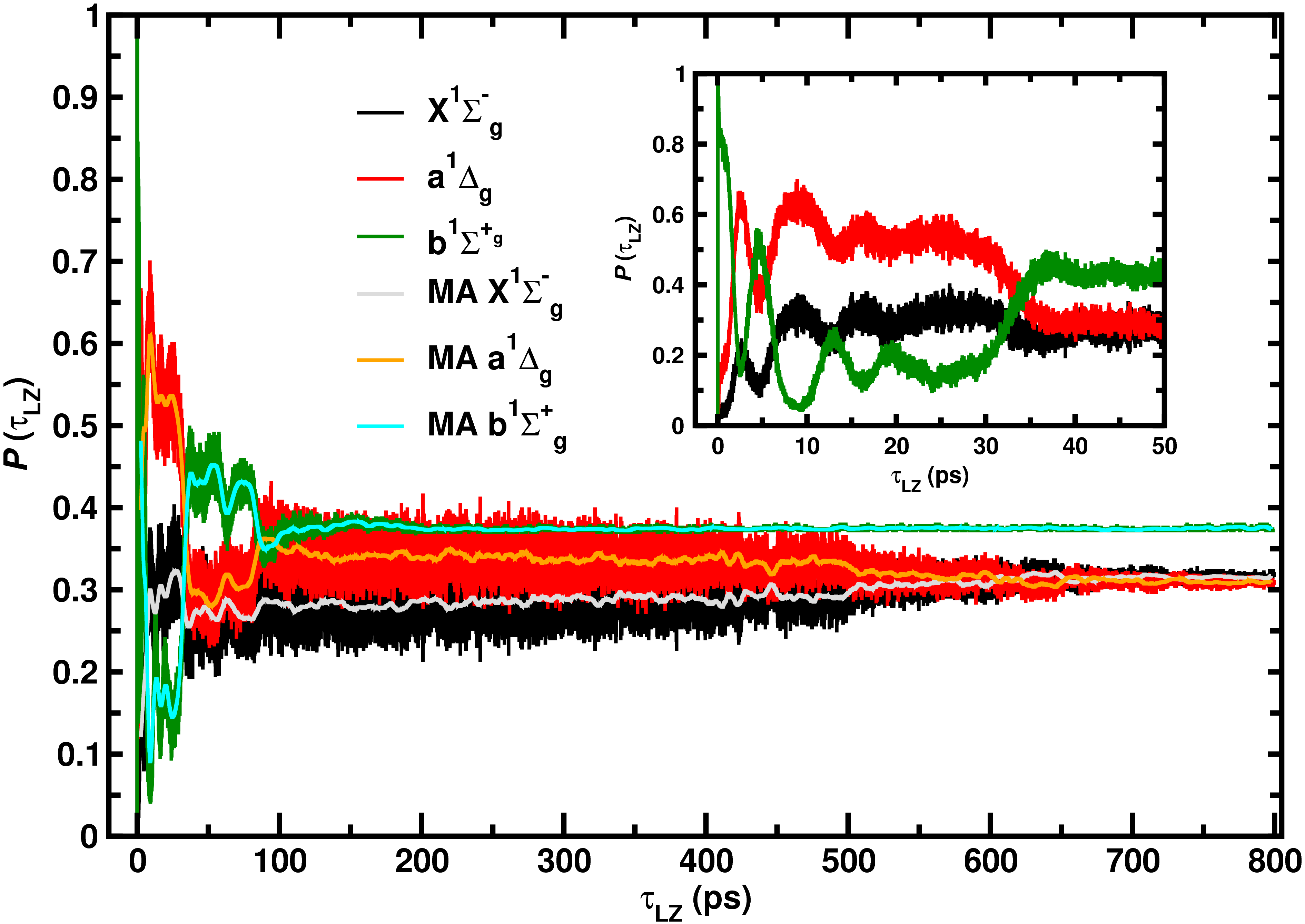}
\caption{Population of the three electronic states and their moving
  average (MA) as a function of time (main panel) and a magnification
  of the first 50 ps in the inset. The zero along the $x-$axis is
  defined as the first transition in each of the 222 trajectories that
  initially rebind into $b^1\Sigma_g^+$. All the population will be
  initially on the $b^1\Sigma_g^+$ state. As the dynamics proceeds,
  the population of the $b^1\Sigma_g^+$ state stabilizes at $\sim 38$
  \% within $\sim 200$ ps. The population dynamics for the
  $X^3\Sigma_g^-$ and $a^1\Delta_g$ states continues until $> 500$
  ps.}
\label{fig:pt}
\end{figure}

\noindent
Transitions $^3\Sigma_g^-$ $\rightarrow$ $a^1\Delta_g$ occur at
smaller interatomic distances (2.09 \AA\/, see Figure S1),
compared to an average value of 2.21 \AA\/ for the $^3\Sigma_g^-$
$\rightarrow$ $b^1\Sigma_g^+$ transition, see Figure
S5B. Including the $^3\Sigma_g^-$ $\rightarrow$
$a^1\Delta_g$ transition into the reaction model leads to an increase
of $\tau_{\rm LZ}$ to $326.8 \pm 226.2$ ps, see Figure
\ref{fig:pt}. While the population on the highest $b^1 \Sigma_{\rm
  g}^+$ state stabilizes after $\sim 200$ ps, transitions between the
$X^3 \Sigma_{\rm g}^+$ and the $a^1 \Delta_{\rm g}^+$ states continue
out to $\sim 700$ ps. Thus, the value of $\tau_c$ represents an
average over the total population distribution for the two states
($^3\Sigma_g^-$ $\rightarrow$ $a^1\Delta_g$, around 500 ps, and
$^3\Sigma_g^-$ $\rightarrow$ $b^1\Sigma_g^+$, 100 ps). Including a
third state with a crossing point at shorter O--O separation makes the
transition time longer. As a consequence, transitions involving
$a^1\Delta_g$ are more prevalent than those involving
$b^1\Sigma_g^+$. The $X^3\Sigma_g^-$ $\rightarrow$ $a^1\Delta_g$
transition occurs with a probability of 76 \% whereas the
$X^3\Sigma_g^-$ $\rightarrow$ $b^1\Sigma_g^+$ occurs for 24 \% with
fluctuations of 22 \%. \\

\noindent
Figure \ref{fig:pt} shows how the population of the three states
differs after the first transition. After recombination on the ground
state, the first transition leads to the $b^1\Sigma_g^+$, common for
all simulations. The first 200 ps are characterized by exchange of
population between the three state. After this time, population on
$b^1\Sigma_g^+$ reaches its equilibrium value. A slower exchange of
population occurs between the two lower states ($^3\Sigma_g^-$ and
$a^1\Delta_g$) for a longer period of time ($\sim 500$ ps) before
reaching the equilibrium population.\\

\noindent
The final state distributions for the three-state model is summarized
in Figure \ref{fig:diagrams}B. For the two-state model, population of
the $b^1\Sigma_g^+$ state is twice more probable than the ground state
$X^3\Sigma_g^-$. For the three-state model, the $b^1\Sigma_g^+$ state
is still most probable, with the $X^3\Sigma_g^-$ and $a^1\Delta_g$
states equally probable. Probabilities for individual events are
comparable with those found for the two states model: the final ground
state population is $\sim 25$ \%, trajectories without collision of
atomic oxygen occurs for $< 20$ \% and the population at the excited
state assume values greater than 50 \%.\\

\begin{figure}
\centering \includegraphics[scale=0.80]{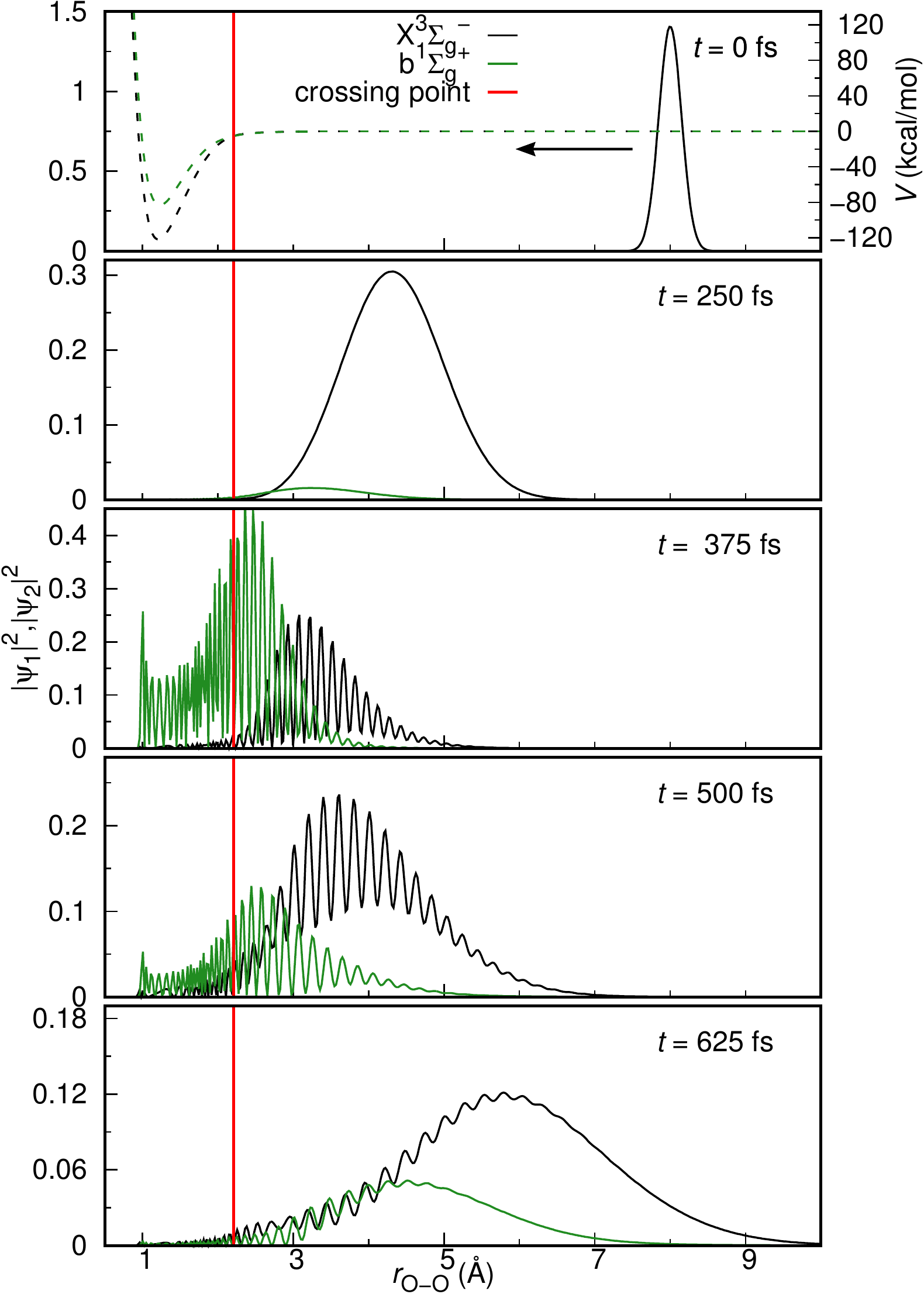}
\caption{Probability distribution of the wave function $|\Psi|^2$
  (solid black and green lines) at different simulation times for both
  states. The potential energies along the diatomic separation are
  also plotted by dashed lines for the corresponding states.  Crossing
  point is shown on `X' axis as red lines. }
\label{fig:qwp}
\end{figure}

\noindent
A final question concerns the validity of classical MD simulations to
follow the population dynamics between two or multiple states. For
this, a wave packet with a translational energy of 2.1 kcal/mol was
initialized on the $X^3\Sigma^-_g$ PES and propagated with a time step
of 0.125 fs allowing for transitions between the $X^3\Sigma_g^-$ and
$b^1\Sigma_g^+$ states. The Fourier transformation of the initial
wavefunction as a function of energy (see Figure S10)
shows that an energy range of 0.03--5.0 kcal/mol is covered which is
required to capture the low-energy part of the motion, characteristic
for a surface temperature of 10 K to 50 K.\\

\noindent
The time evolution of the wave function is shown in Figure
\ref{fig:qwp} (black for $X^3\Sigma_g^-$ and green for
$b^1\Sigma_g^+$). Initially ($t=0$, $r=8.0$ \AA\/), the system is on
the $X^3\Sigma_g^-$ surface.  The radial grid consists of 1250 evenly
spaced points from 0.35 to 26.8 \AA\ and the damping function starts
at 15.0 \AA. After $\sim 250$ fs the wavepacket has passed the
crossing region, splits into two parts, and the amplitude on the
$b^1\Sigma^+_g$ PES starts to increase.  The distribution of
population as a function of time is shown in Figure
S9. Significant amount of population transfers from the
$X^3\Sigma^-_g$ state to the $b^1\Sigma^+_g$ state occurs up to 380
fs, although some amount of the WP transfers back to the
$X^3\Sigma^-_g$ state due to the coupling of the two states. Since the
QM simulations are carried out in the gas phase, only one single
collision can be followed. However, in the condensed phase,
vibrational relaxation can form a bound O$_2$ molecule. Multiple
crossings are thus possible for the high lying vibrational states of
O$_2$ molecule.\\

\noindent
To account for the recurrences, a kinetic model has been constructed
for multiple crossings of the WP, see Figure S11.  In gas
phase the transition probabilities from one state to another starting
from any of the states are equal. Hence, the kinetic model for 2
states leads to a stationary population of 50\% on each of the
$X^3\Sigma^-_g$ and $b^1\Sigma^+_g$ states. Here, it is worth
mentioning that this ratio is 66 \% on $b^1\Sigma^+_g$ and 34 \% on
$X^3\Sigma^-_g$ obtained from the condensed phase classical MD
simulations. A possible explanation for this observation is the
different coupling between the O$_2$ motion and the surrounding water
matrix due to the different curvatures of the potential energy curves
for the two electronic states. A similar kinetic model for 3-state
model leads to a stationary statistical population of $1/3$ on each
state, which is close to the classical MD simulations (31 \%, 31 \%,
and 38 \% for the $X^3\Sigma^-_g$, $a^1\Delta_g$ and $b^1\Sigma^+_g$
states, respectively). \\

\section{Conclusion}
The present work establishes that upon recombination of $^3$P atomic
oxygen on ASW, molecular oxygen (O$_2$) in its ground and lower
electronically excited states can be formed and stabilized. It should
be emphasized that desoprtion of O$_2$ after formation through
association of atomic oxygen, although energetically feasible, was not
observed. This is consistent with earlier work.\cite{MM.oxy:2019}

\noindent
For singlet oxygen, the radiative decay lifetimes have been determined
in the gas phase. They range from $\sim 1$ min to $\sim 100$ min for
the $a ^1\Delta_g$ and $b ^1 \Sigma_g$ states and also depend on the
vibrational level.\cite{rothman:1998,ballard:1999} Thus, the actual
fraction available for chemical processes will depend on how the
radiative lifetime changes when O$_2$ is adsorbed on ASW. Because the
reactivity of the $a^1\Delta_g$ and $b^1\Sigma_g^+$ states can be
considerably larger than that of the $X^3\Sigma_g^-$ ground state
depending on the reaction partner, including electronically excited
states of O$_2$ (and other molecules formed on ASW) may be essential
for a comprehensive modeling and understanding of the chemistry of
interstellar matter under such conditions.\\

\section*{Acknowledgments}
This work was supported by the Swiss National Science Foundation
through grants 200021-117810, and the NCCR MUST.

\bibliography{o2}{} 
\bibliographystyle{plain}

\end{document}


\section{Methods}
The simulation system consists of an equilibrated cubic box of
amorphous solid water with dimension $31 \times 31 \times 31$
\AA\/$^3$ containing 1000 TIP3P\cite{jorgensen-tip3p} water molecules,
and two oxygen atoms, see Figure 1. The time step in all
simulations was $\Delta t =0.1$ fs which ensures conservation of total
energy also during the recombination dynamics. All bonds involving
hydrogen atoms were constrained using SHAKE\cite{shake} and the
non-bonded cutoff was at 13 \AA\/. Initial conditions were generated
from an existing, equilibrated ASW
structure\cite{MM.oxy:2018,MM.oxy:2019} by adding two oxygen atoms,
minimizing the system, heating it for 5 ps to 50 K and equilibrating
for 10 ps. Data (energies, coordinates and velocities) are saved every
1,000 steps. This was followed by production simulations of various
lengths, as indicated throughout this work. \\

\begin{figure}
\centering \includegraphics[scale=0.60]{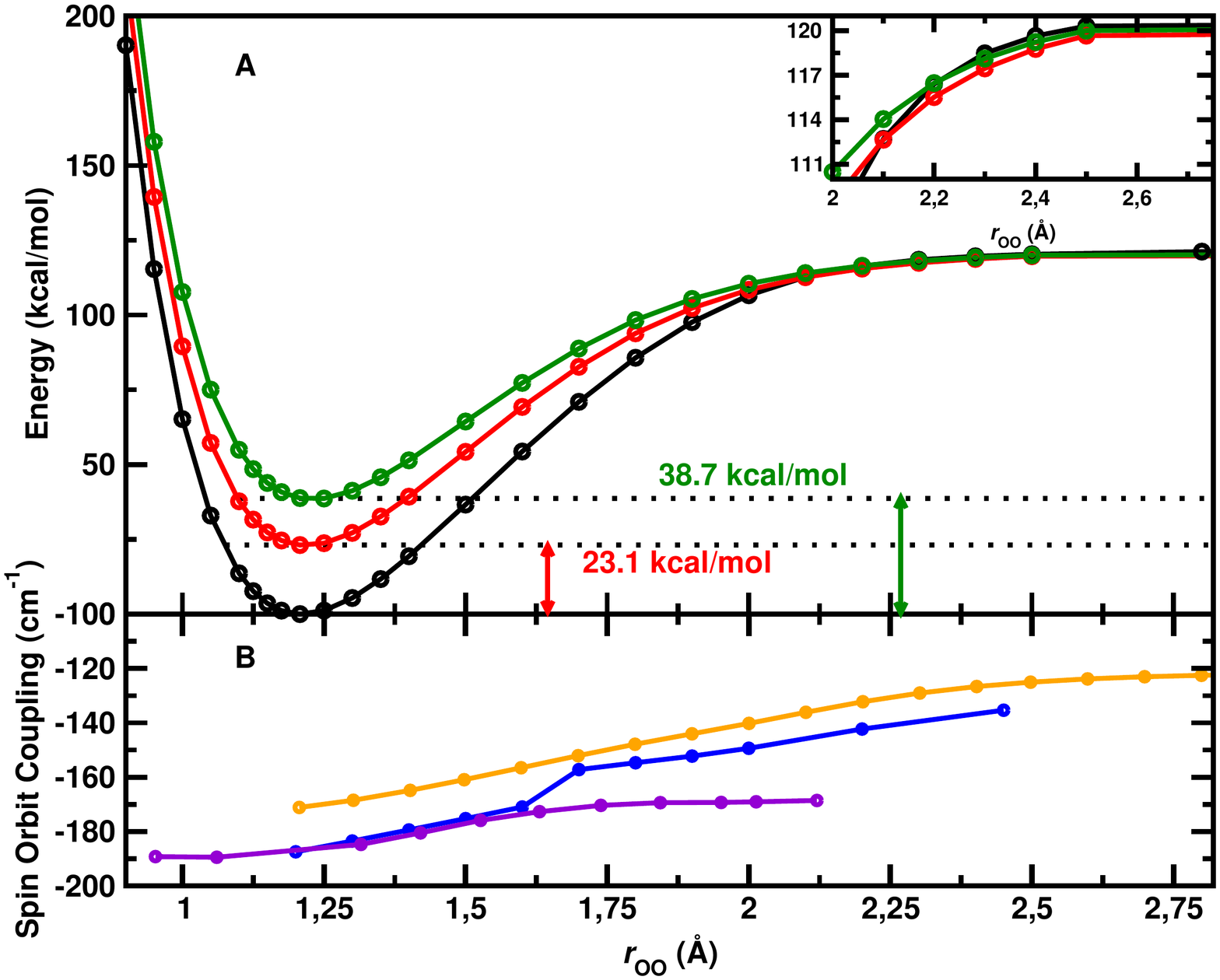}
\caption{Panel A: Potential energy curves for the $X ^3\Sigma_g^-$, $a
  ^1\Delta_g$ and the $b ^1\Sigma_g^+$ states (black, red and green,
  respectively) with the crossing region enlarged in the inset. Panel
  B: SOC between the $^3\Sigma_g^-$ and $^1\Sigma_g^+$ states from the
  literature (blue\cite{minaev2003} and violet\cite{dayou2005}) and
  the SOC calculated in the present work (orange). The differences can
  be explained by the difference in methodologies and basis set
  between used in the different calculations. Asymptotically, the SOC
  approaches twice the value of O $^3P$, which is 74.182 cm$^{-1}$ for
  a single oxygen atom in its ground state.}
\label{fig:fig2}
\end{figure}

\noindent
All simulations are performed using the CHARMM
program\cite{charmm.prog} modified for reactive MD
simulations\cite{mmrev:2018,msarmd} and potential energy surfaces for
O$_2$ based on reproducing
kernels.\cite{hollebeek1999constructing,MM.rkhs:2017} For treating
nonadiabatic transitions a surface hopping scheme based on the Landau
Zener formalism is used, see below. Three electronic states for O$_2$
are considered: the ground state ($X ^3\Sigma_g^-$) and the next two
electronically excited states ($a ^1\Delta_g$ and $b ^1 \Sigma_g^+$)
based on earlier calculations at the MRCI/aug-cc-pVTZ level of
theory\cite{Ruede:2010} which were accurately represented as a
reproducing kernel Hilbert space (see Figure
\ref{fig:fig2}a).\cite{hollebeek1999constructing,MM.rkhs:2017} The
spin orbit coupling (SOC) matrix elements involving the $^3
\Sigma_{\rm g}^-$ and $^1 \Sigma_{\rm g}^+$ states were recomputed at
the MRCI/aug-cc-pVTZ level of theory and can be compared with previous
works at the CI/cc-pVTZ\cite{minaev2003} and CASSCF/CASPT2/$5s4p3d2f$
atomic natural orbital basis set\cite{dayou2005} levels of theory, see
Figure \ref{fig:fig2}b. Because the $a ^1\Delta_g$ and the $b
^1\Sigma_g^+$ states do not cross (see inset Figure \ref{fig:fig2})
and the nonadiabatic coupling (NAC) matrix element between the two
states is zero, transitions between these two states will not be
included.\\

\noindent
In the gas phase all transitions between the $^3\Sigma_g^-$,
$^1\Delta_g$ and $^1\Sigma_g^+$ states are strictly
forbidden. Transitions from the ground state to the two excited states
are spin-forbidden\cite{herzberg2} and Laporte rule\cite{Laporte:25},
because all states have $g$ symmetry. Due to the second rule,
transitions between $a^1\Delta_g$ and $b^1\Sigma_g^+$ are also
forbidden. However, because the reaction occurs on the ASW surface, no
symmetry rules apply for allowed transitions. This is similar to the
fact that $Q-$branches for diatomics in liquids and high pressure
fluids become allowed due to symmetry breaking induced by the
environment,\cite{vodar:1960} whereas such $\Delta J = 0$ transitions
are forbidden in free space.\cite{herzberg}\\

\noindent
For the nonadiabatic transitions the trajectory surface hopping (TSH)
method\cite{sti76:3975} within the Landau-Zener
(LZ)\cite{lan32:46,zen32:696} formalism was used. The implementation
follows earlier work\cite{bel11:014701,bel14:224108} for which the
transition probability $P_{\rm LZ}^{i\rightarrow j}$ from state $j$ to
$k$ is
\begin{equation}
P_{\rm LZ}^{j\rightarrow k} = exp \left( -2 \pi \frac{\left(\Delta
  H_{jk} \right)^2 }{\hbar \vert \dot{\vec{R}} \cdot
  \vec{\nabla}(\Delta E_{jk}) \vert} \right)
\end{equation}
The transition probability depends on the gradient of the energy
difference $\vec{\nabla}(\Delta E_{jk})$ between states $j$ and $k$,
the coupling $\Delta H_{jk}$, which is the conformationally dependent
spin orbit matrix element, and the velocity of the center of mass
$\dot{\vec{R}}$ at the transition.\\

\noindent
The trajectories are started from a given initial electronic state $j$
and the electronic state is followed along the trajectory. Close to a
crossing between the present state $j$ and a neighboring state $k$,
$P_{\rm LZ}^{j\rightarrow k}$ is calculated and compared with a random
number $\xi \in [0,1]$. If $P_{\rm LZ}^{j\rightarrow k} \geq \xi$ a
transition from state $j$ to state $k$ occurs. To ensure conservation
of the total energy and total angular momentum, a momentum correction
\begin{equation}
 \bf{p'} = p - \hat{n}\frac{\hat{n}M^{-1}p}{\hat{n}M^{-1}\hat{n}}\left
    [{\rm 1}-\left({\rm 1-2\Delta E}
      \frac{\hat{n}M^{-1}\hat{n}}{(\hat{n}M^{-1}p)^{\rm 2}} \right
      )^{\rm 1/2}\right ],
\end{equation}
is applied\cite{mil72:5637} where $\mathbf{p}$ and $\mathbf{p'}$ are
the momenta before and after the hop and $\mathbf{M}$ is the mass
matrix and $\hat{\mathbf{n}}$ is the unit vector along the velocity
direction.\\

\noindent
In order to assess the role of quantum effects on the nuclear
dynamics, time-dependent wave packet (WP) simulations were carried out
for the two state model (see below), including the $X^3\Sigma_g^-$ and
$b^1\Sigma_g^+$ states. For this, the time-dependent Schr\"{o}dinger
equation is\cite{pru03:2354,mah01:2321,kop07}
\begin{equation}
\begin{pmatrix} 
\psi_1(r;t) \\
\psi_2(r;t) \\
\end{pmatrix} =e^{-iHt/\hbar} \begin{pmatrix} 
\psi_1(r;0) \\
\psi_2(r;0) \\
\end{pmatrix}
\end{equation}
where the Hamiltonian $H$ is
\begin{equation}
\hat{H} =-\frac{\hbar^2}{2\mu}\frac{\partial^2}{\partial r^2}
+ \begin{pmatrix} V_{11}(r) & V_{12}(r) \\ V_{21}(r) & V_{22}(r) \\
\end{pmatrix}
\end{equation}
where $\mu$ is the reduced mass of the system, $V_{11}$, $V_{22}$ are
the diabatic potential energies and $V_{12} = V_{21}$ is the
geometry-dependent coupling matrix element between two states.\\

\noindent
The initial wave packet is a (complex-valued) Gaussian function
\begin{equation}
\psi_0(r) = (1/2\pi\sigma^2)^{1/4} \exp[-1/(2\sigma^2)(r-r_0)^2) \exp[ip_0(r-r_0)],
\end{equation}
where $\sigma$ is the width parameter, $r_0$ and $p_0$ are the initial
position and momentum of the wavepacket, respectively. The
time-dependent wave function is propagated\cite{kon13:13070} on the
coupled $X^3\Sigma_g^+$ and $b^1\Sigma_g^+$ potentials using the
split-operator method.\cite{fei82:412} Fast Fourier transformation
(FFT) is used to calculate the double differentiation
$\frac{\partial^2}{\partial r^2}$ of the wave function. A sine damping
function is multiplied to the wave function at the grid boundary to
avoid reflection. The state population can then be calculated as the
expectation value of the projection operator\cite{kop07}
\begin{equation}
P_2(t) =\left \langle \begin{pmatrix} \psi_1(x;t) \\ \psi_2(x;t) \\
\end{pmatrix} \left | \begin{pmatrix} 
0 & 0 \\
0 & 1 \\
\end{pmatrix}
\right |  \begin{pmatrix} 
\psi_1(x;t) \\
\psi_2(x;t) \\
\end{pmatrix} \right \rangle
\end{equation}
and $P_1(t) = 1 - P_2(t)$.\\

\begin{figure}
\centering
\includegraphics[scale=0.4]{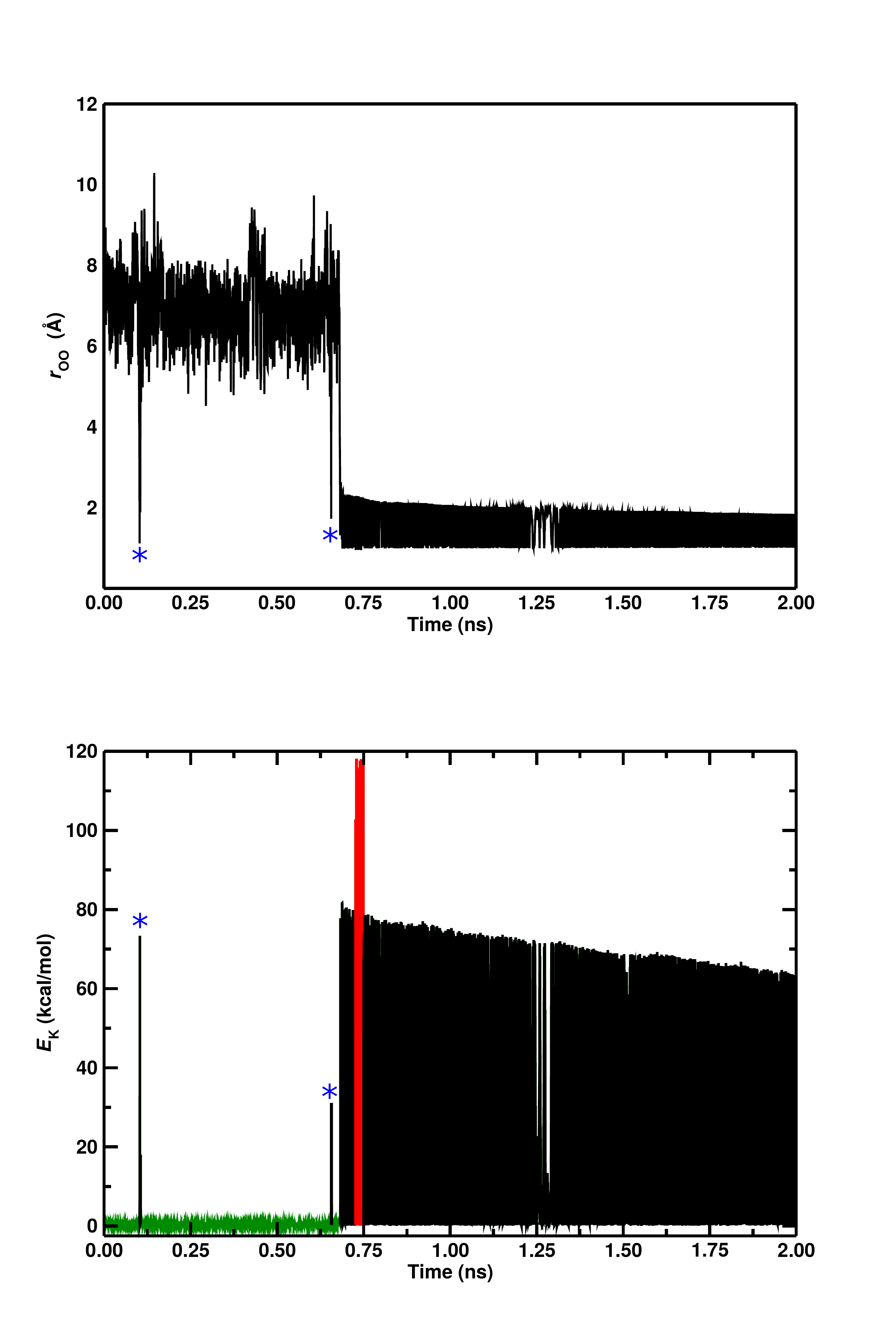} 
\caption{Top panel: time evolution of the interatomic distance between
  two oxygen atoms for the simulation inside the cavity. The first
  time that the bound state is reached, at 103.5 ps, it suddenly
  dissociates due the collision of the molecule with one of the TIP3P
  hydrogen water. A similar situation happens near the formation time
  (around 670 ps). Both are shown in the graph by a blue star. Bottom
  panel: kinetic energy for the dioxygen system. The bound state can
  be recognized by the peak in kinetic energy at 103.5 ps and at 670
  ps (shown with a blue star). The transition between $^1\Sigma_g^+$
  and $^3\Sigma_g^+$ is observed between 720 ps and 750 ps and is
  characterized by the $\sim$40 kcal/mol energy increase. This energy
  corresponds to the energy difference between the two states. A color
  code here is applied in order to distinguish the three different
  states: the green line represents the unbound state, the red one the
  time interval where the triplet region is explored and the black is
  used when O$_2$ lays in the singlet state.}
\label{sifig:inside}
\end{figure}

\begin{figure}
\centering \includegraphics[scale=0.6]{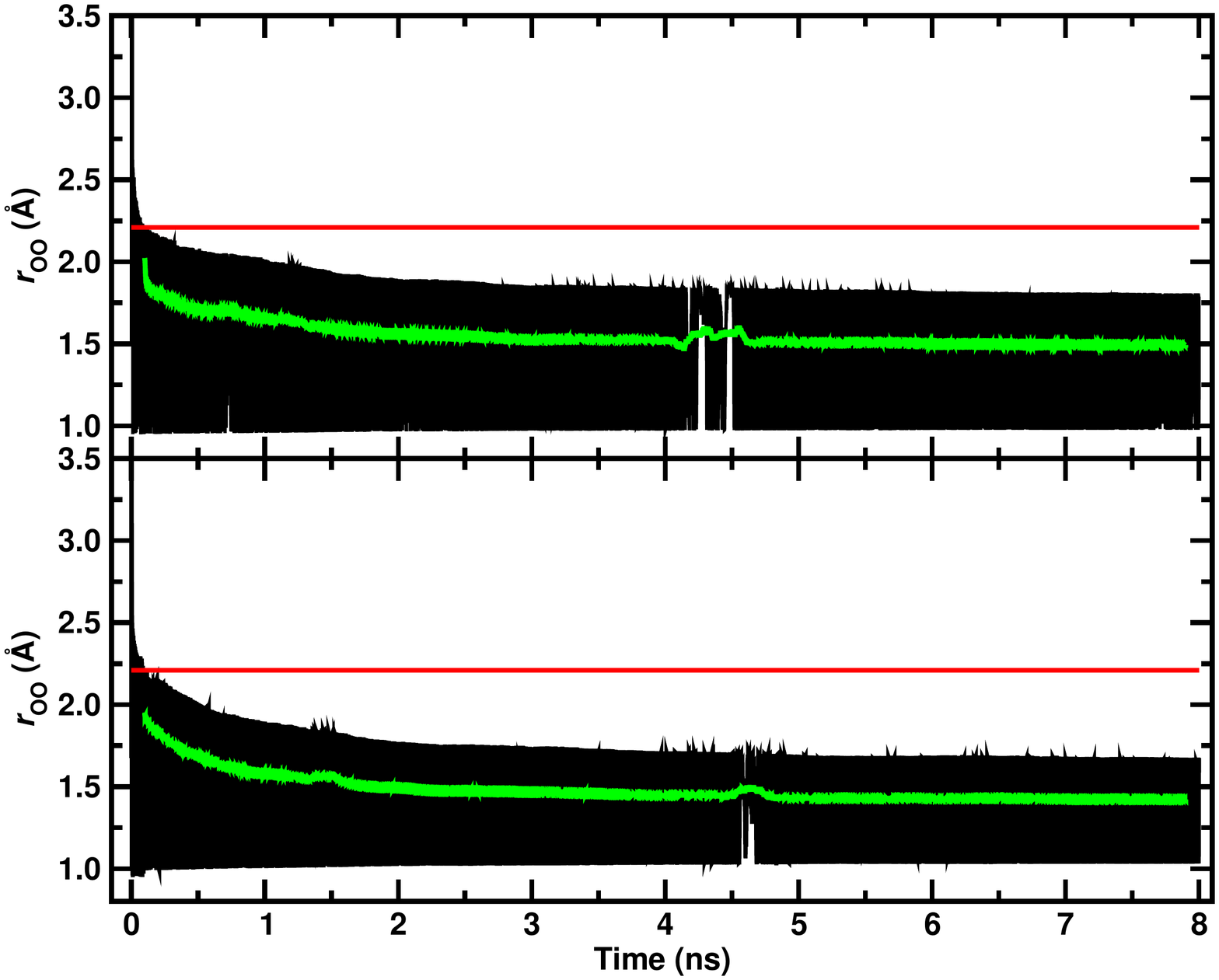}
\caption{Relaxation of the O$_2$ bond length during two independent 8
  ns simulation with the two states models for two
  simulations. Transitions are observed only in the initial steps of
  the simulations corresponding to the overlap between the time series
  and the crossing point between the two states (2.209 \AA\/, red
  line). The green line represents the moving average of the time
  series over 0.2 ps time interval. The signatures between 4 ns and
  4.5 ns are collisional re-excitation of the diatomic due to
  collisions with the surface.}
\label{sifig:relax8ns}
\end{figure}

\begin{figure}
\centering
\includegraphics[scale=0.60]{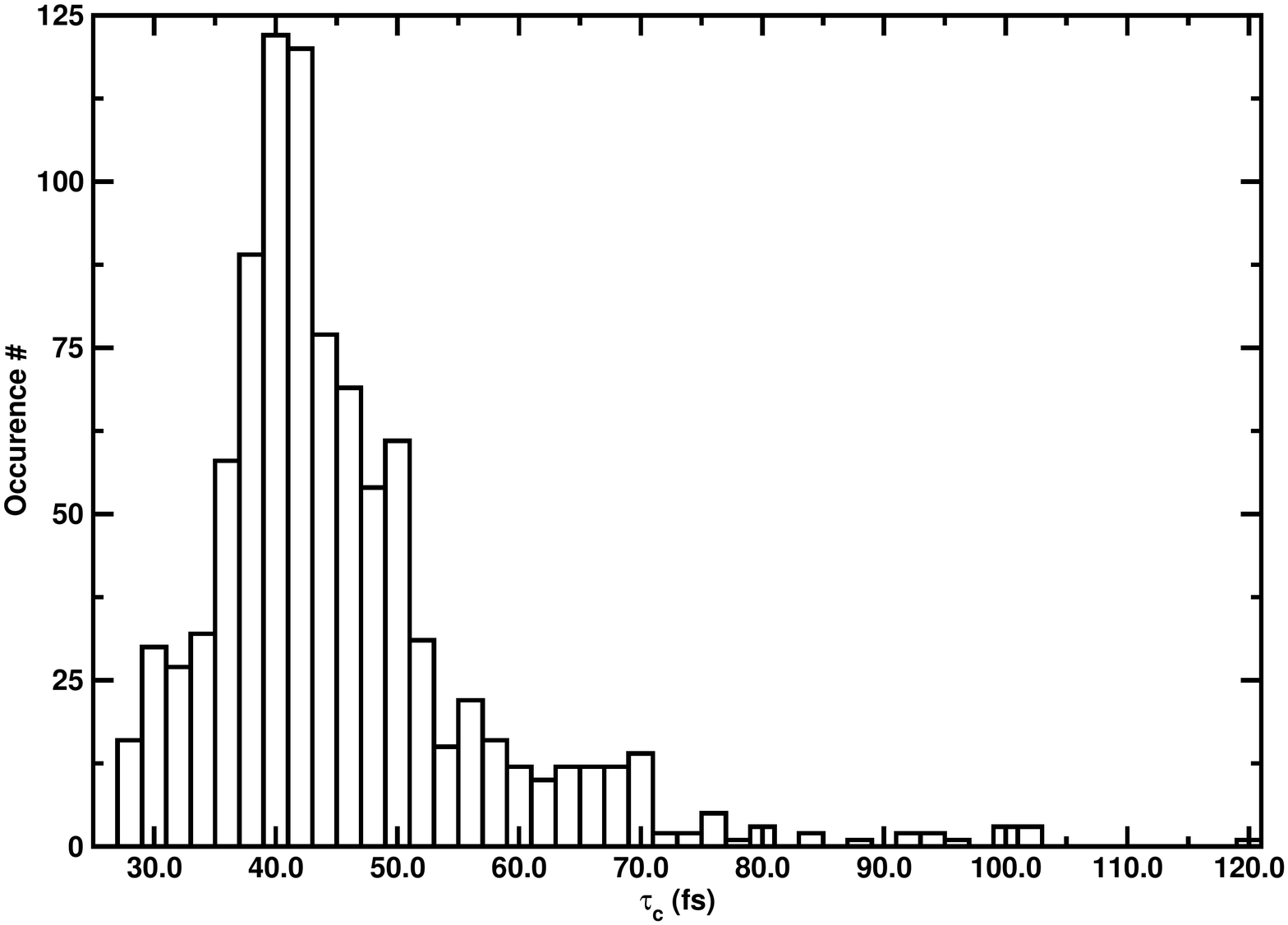} 
\caption{Probability distribution for $\tau_{\rm c}$. The average is
  $47.4 \pm11.7$ fs. This corresponds to approximately one transition
  every two vibrational periods.}
\label{sifig:tttrans}
\end{figure}

\begin{figure}
\centering \includegraphics[scale=0.60]{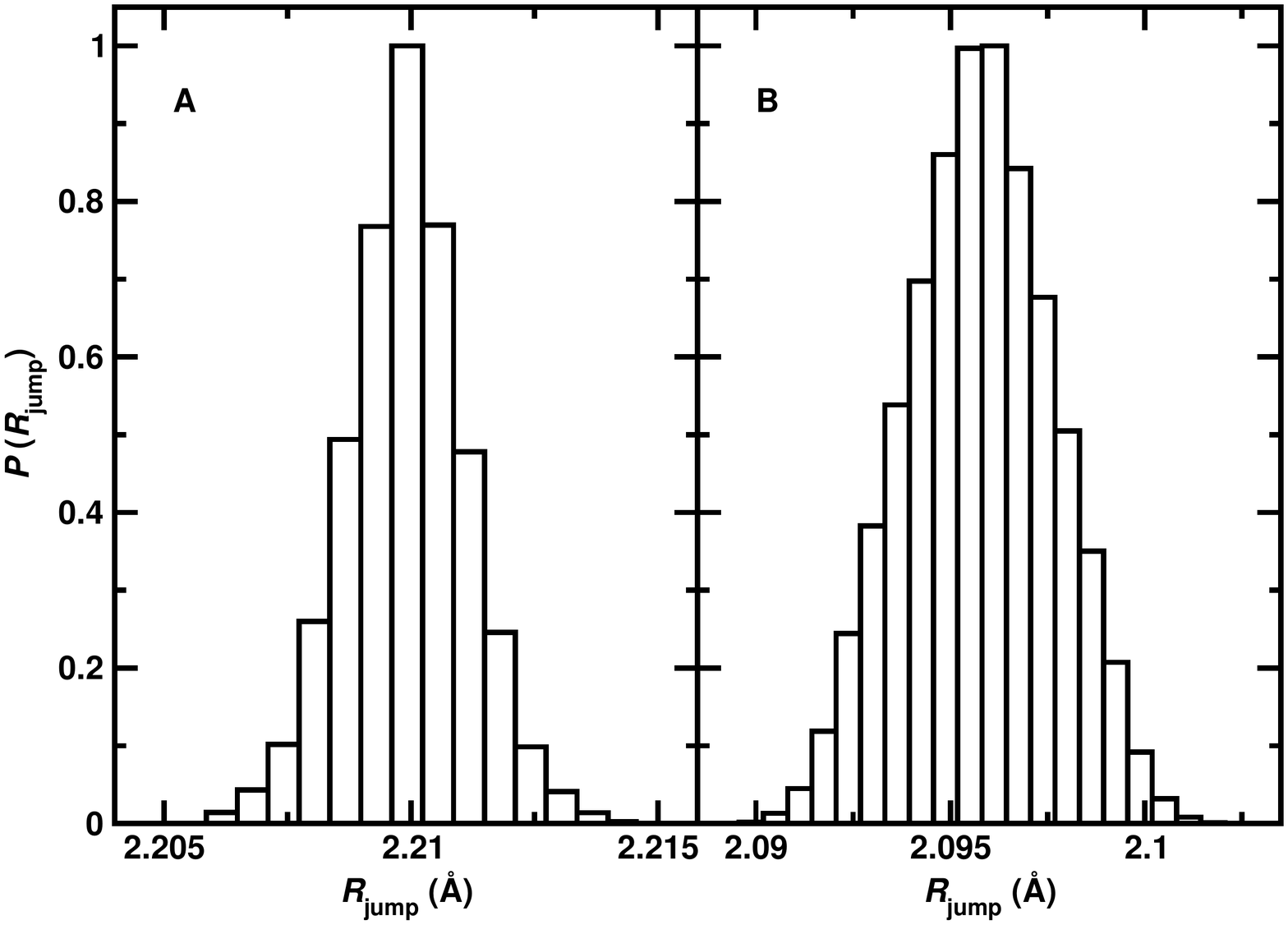}
\caption{Probability distributions of the
  interatomic distance within the transition between the
  $a^1\Delta_g^-$ and the $X^3\Sigma_g^-$ (B), b$^3\Sigma_g^-$ and
  the $X^3\Sigma_g^-$ (A) for an ensemble of simulations.. In the
  first case the interval is localized around 2.097 $\pm$ 0.003 \AA\/
  and in the second case around 2.209 $\pm$ 0.003 \AA\/.}
\label{sifig:rjump}
\end{figure}

\begin{figure}
\centering \includegraphics[scale=0.60]{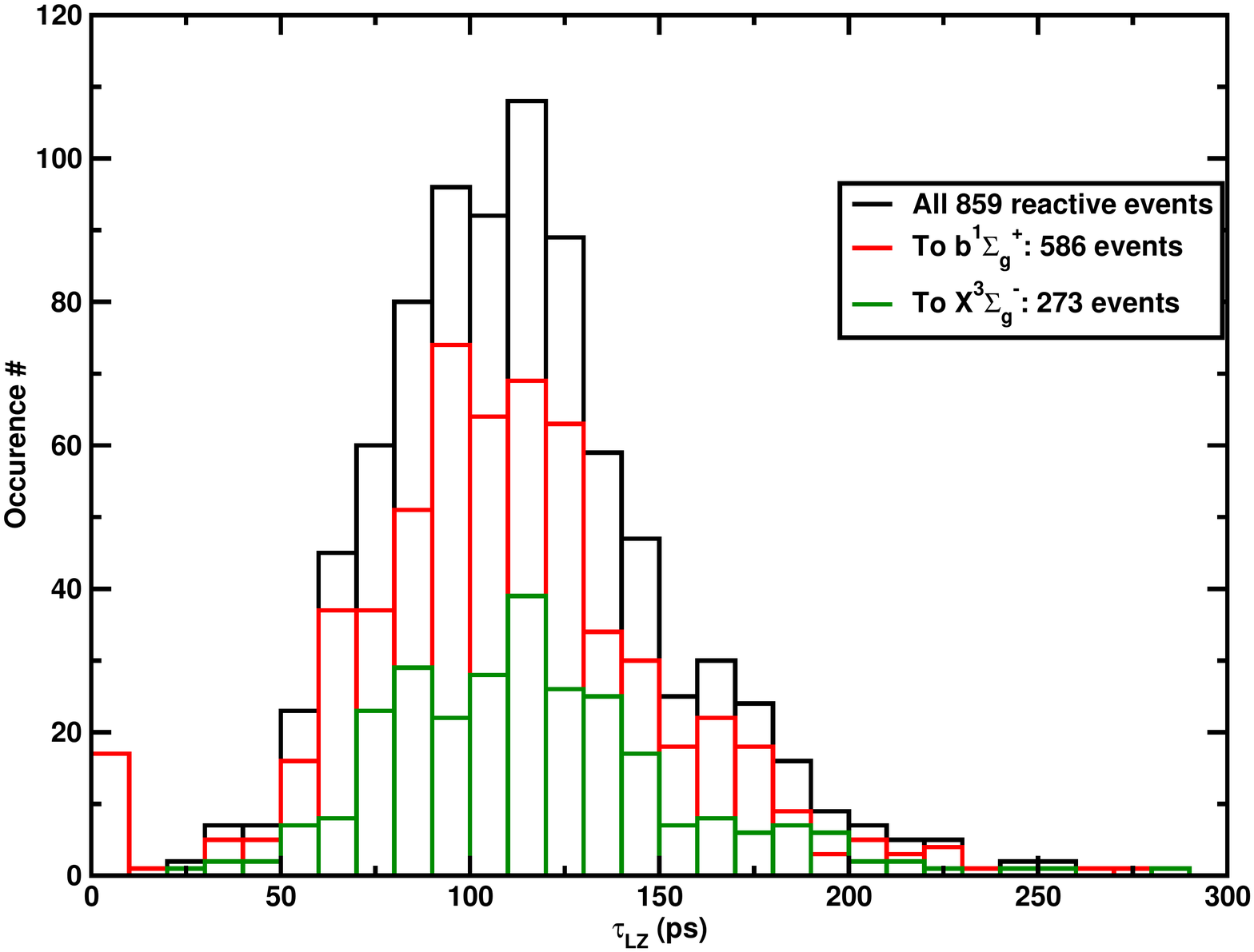}
\caption{The time distribution for the three $\tau_{LZ}$ for the O$_2$
  recombination from b$^3\Sigma_g^-$, in red simulations that from the
  initial $X^3\Sigma_g^-$state leads to final $b^1\Sigma_g^+$ state,
  in green simulations that have $X^3\Sigma_g^-$ as initial and final
  state The and in black the sum of the two previous sets.}
\label{sifig:hist_tz_s2}
\end{figure}

 \begin{figure}
\centering
\includegraphics[scale=0.60]{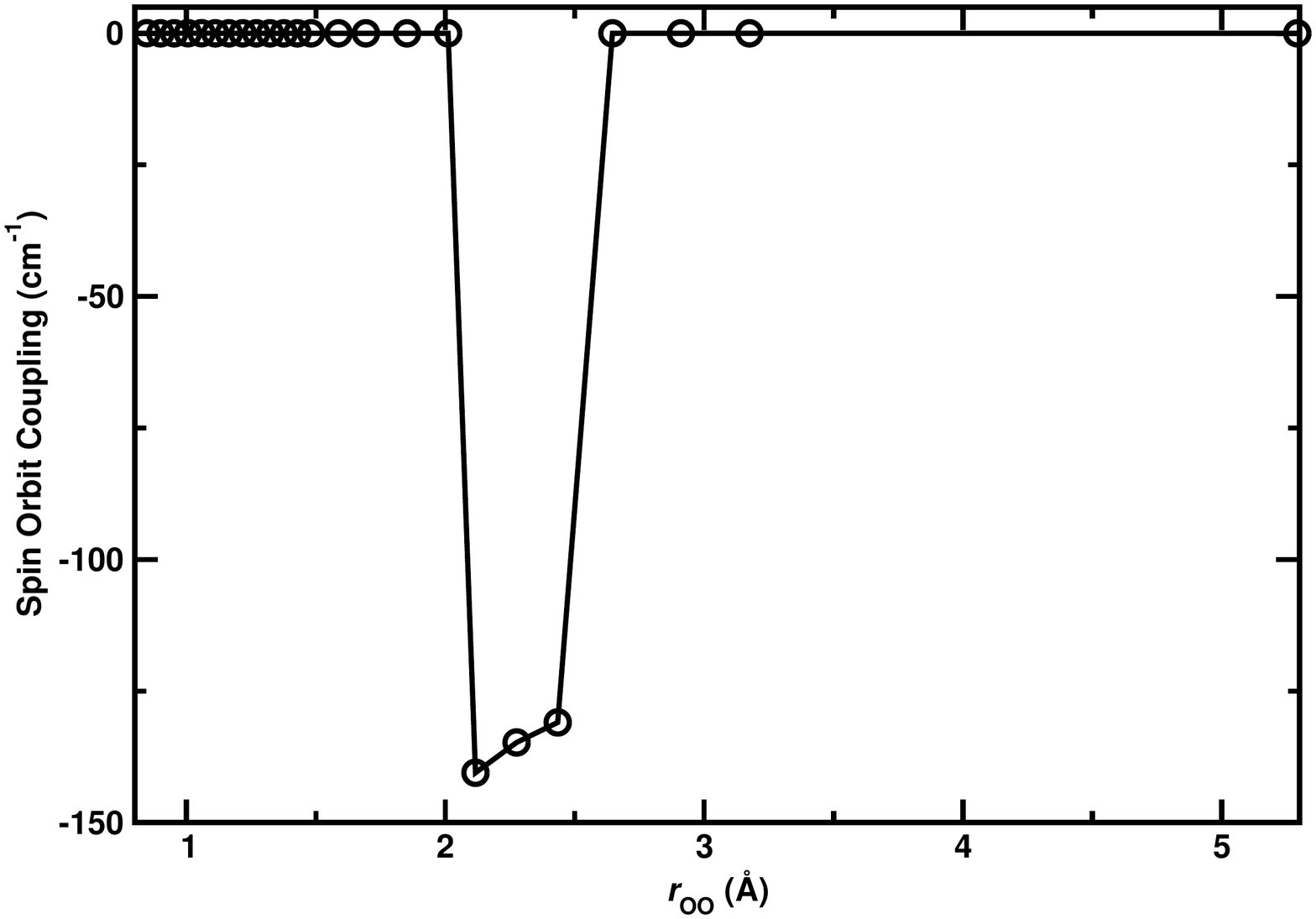} 
\caption{Computed spin orbit coupling between $X^3\Sigma_g^-$and
  $a^1\Delta_g$. The function is discontinuous, with values different
  from 0, only in the crossing regions. The two atoms are in the $^3$P
  state.}
\label{sifig:s1soc}
\end{figure}

\begin{figure}
\centering \includegraphics[scale=0.60]{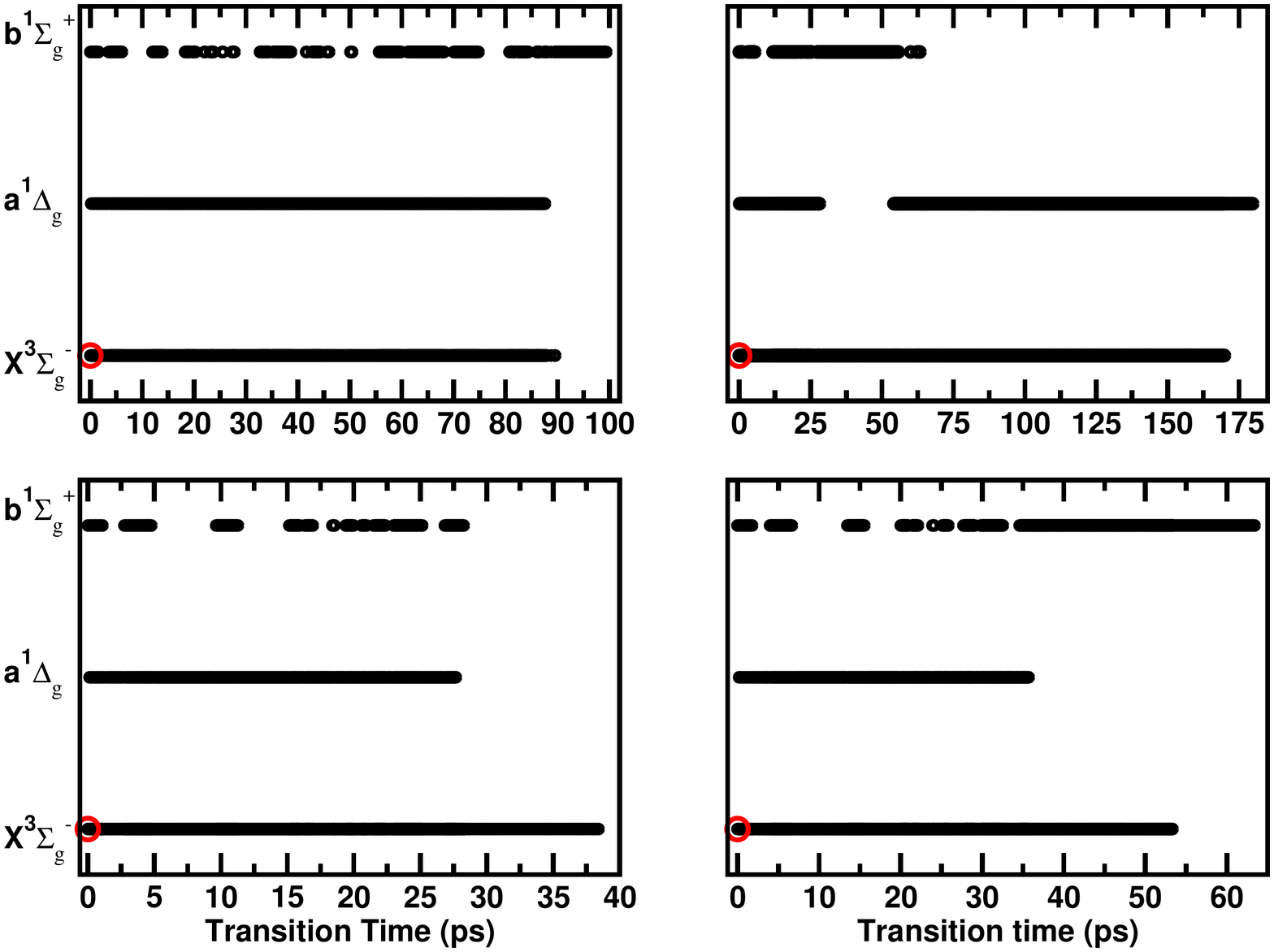}
\caption{State dynamics for four different simulations during the
  $\tau_{\rm LZ}$ interval. All simulations start from the
  $X^3\Sigma_g^-$ state (red dot).}
\label{sifig:threetrans}
\end{figure}

\begin{figure}
\centering \includegraphics[scale=0.80]{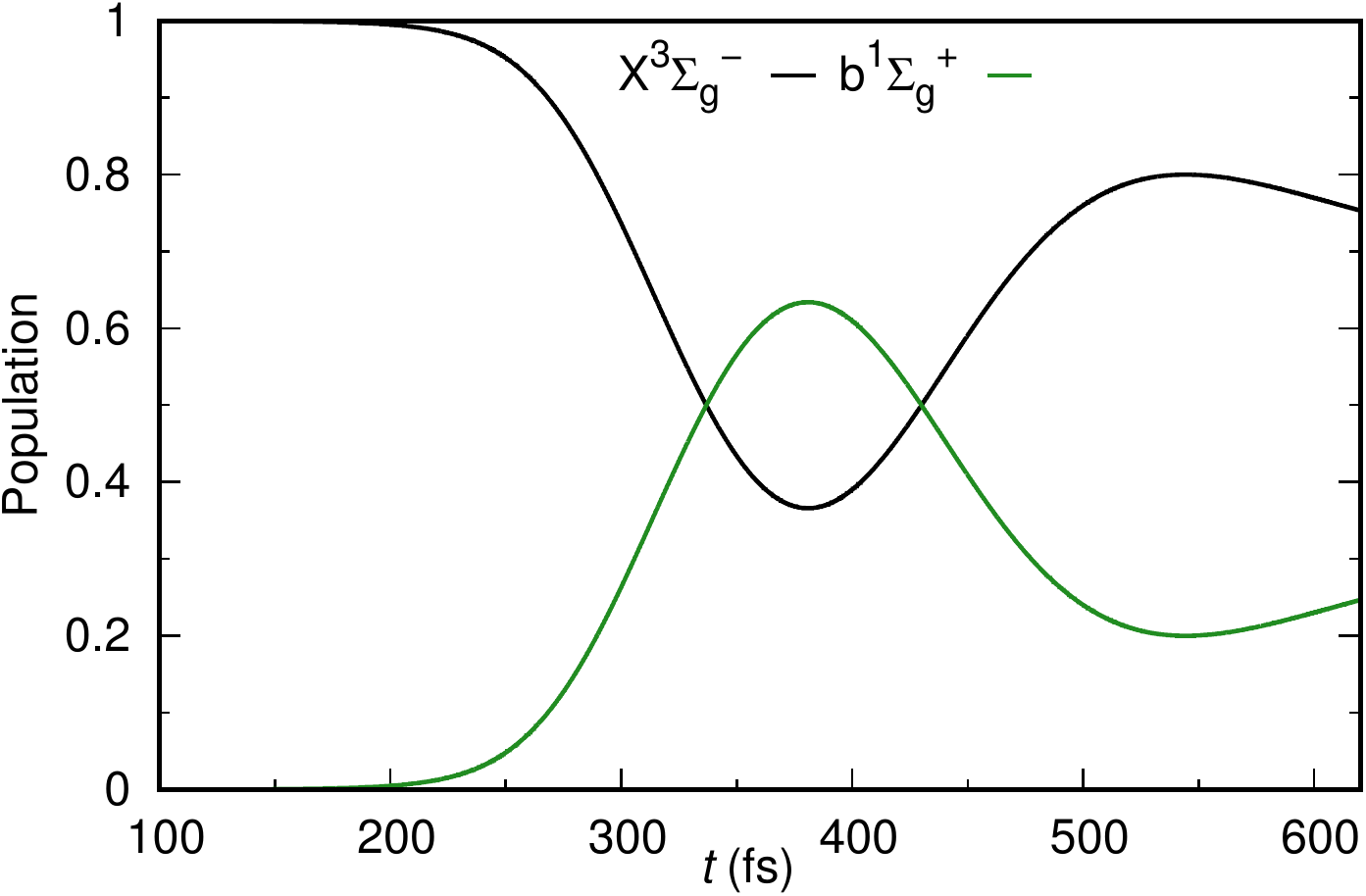}
\caption{Population on each state as a function of time.}
\label{fig:qpop}
\end{figure}

\begin{figure}
\centering \includegraphics[scale=0.70]{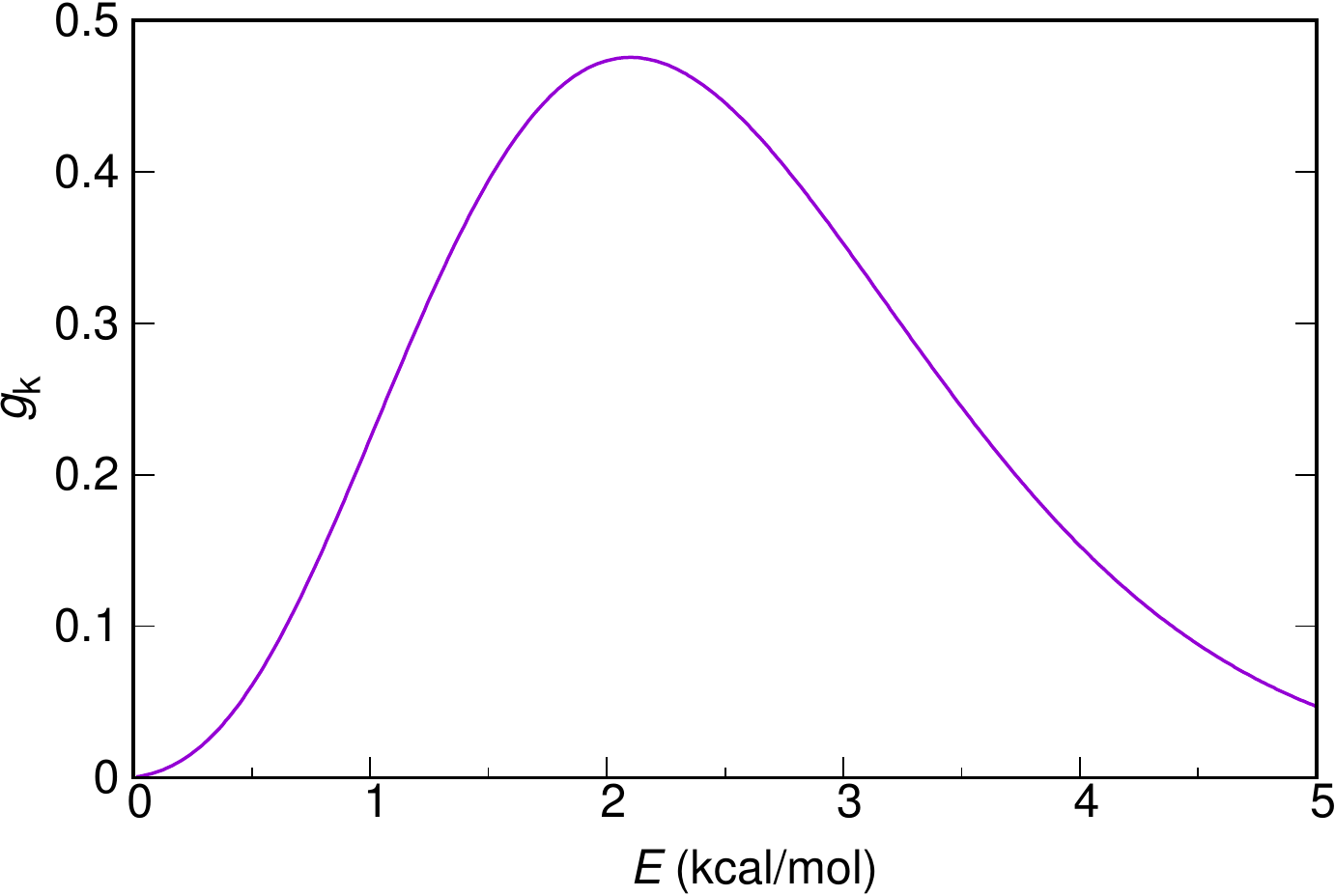}
\caption{Fourier transform of the initial wave packet on $X^3\Sigma_g^-$
state as a function of energy. The initial wave packet is a Gaussian function
defined in Eq. 5 and  $g(k) = \frac{\sqrt{2\sigma}}{(2\pi)^{1/4}}$exp$[-\sigma^2(k-p_0)^2)$exp$(ir_0k)$.
where, $k=\sqrt{2\mu E}$ and $p_0 = \sqrt{2\mu E_0}$.}
\label{sifig:edist}
\end{figure}

\begin{figure}
  \centering
  \includegraphics[scale=0.70]{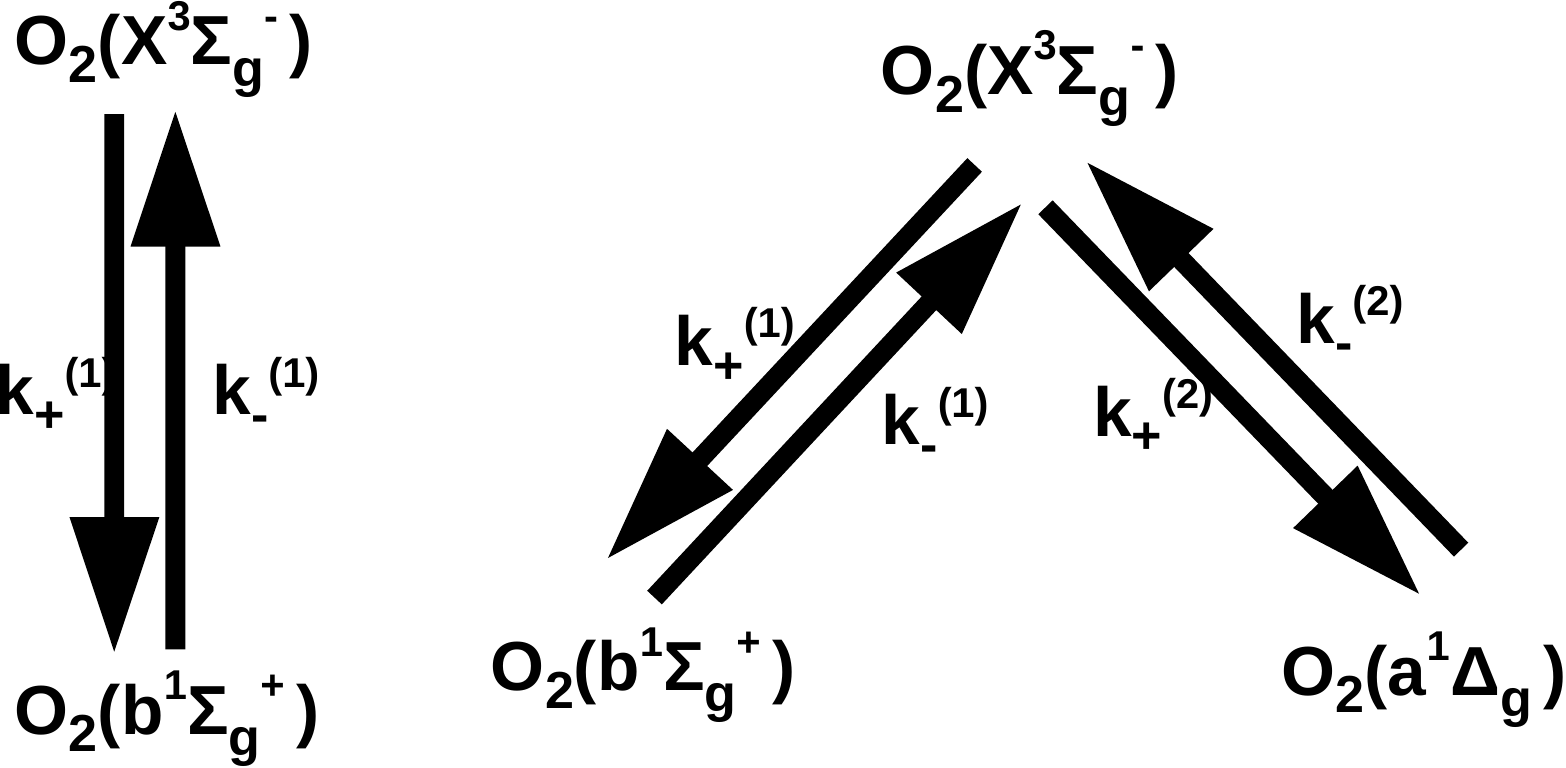}
\caption{Kinetic model for multiple crossing. Left: 2-state model,
  right: 3-state model.  Here, $k$s are the probabilities for the
  transition from one state to another.}
\label{sifig:km}
\end{figure}

\clearpage
\newpage
\bibliography{o2}{} 
\bibliographystyle{plain}